# Disentangling coincident cell events using deep transfer learning and compressive sensing

*Moritz Leuthner\*, Rafael Vorländer, Oliver Hayden\**

M. Leuthner[1,2,3], R. Vorländer[1,3], O. Hayden[1,2,3]
[1] Heinz-Nixdorf-Chair of Biomedical Electronics, School of Computation, Information and Technology, Technical University of Munich, Einsteinstraße 25, 81675 Munich, Germany
[2] Munich Institute of Biomedical Engineering, School of Computation, Information and Technology, Technical University of Munich, Boltzmannstraße 11, 85748 Garching b. Munich, Germany
[3] Central Institute for Translational Cancer Research (TranslaTUM), University Hospital Rechts der Isar, Technical University of Munich, Einsteinstraße 25, 81675 Munich, Germany
E-mail: moritz.leuthner@tum.de; oliver.hayden@tum.de

**Abstract**

Accurate single-cell analysis is critical for diagnostics, immunomonitoring, and cell therapy, but coincident events—where multiple cells overlap in a sensing zone—can severely compromise signal fidelity. We present a hybrid framework combining a fully convolutional neural network (FCN) with compressive sensing (CS) to disentangle such overlapping events in one-dimensional sensor data. The FCN, trained on bead-derived datasets, accurately estimates coincident event counts and generalizes to immunomagnetically labeled CD4 and CD14 cells in whole blood without retraining. Using this count, the CS module reconstructs individual signal components with high fidelity, enabling precise recovery of single-cell features, including velocity, amplitude, and hydrodynamic diameter. Benchmarking against conventional state-machine algorithms shows superior performance—recovering up to 21% more events and improving classification accuracy beyond 97%. Explainability *via* class activation maps and parameterized Gaussian template fitting ensures transparency and clinical interpretability. Demonstrated with magnetic flow cytometry (MFC), the framework is compatible with other waveform-generating modalities, including impedance cytometry, nanopore, and resistive pulse sensing. This work lays the foundation for next-generation non-optical single-cell sensing platforms that are automated, generalizable, and capable of resolving overlapping events, broadening the utility of cytometry in translational medicine and precision diagnostics, e.g. cell-interaction studies.

## 1. Introduction

Single-cell analysis is a cornerstone of modern biomedical research[1], underpinning advances in diagnostics[2,3], immunophenotyping[4-6], cell therapy monitoring[7-9], and systems biology[10-12]. High-throughput cytometric platforms—including optical flow cytometry[13-15], Coulter-type impedance analyzers[16-19], and magnetic flow cytometry (MFC)[20-22]—enable precise identification and quantification of cellular subpopulations based on physical, biochemical, or marker-based properties. Despite their differing sensing modalities, these platforms share a common principle: the generation of quasi-linear, one-dimensional time-series signals as cells transit confined sensing regions.

In optical flow cytometry, time-resolved fluorescence or scatter signals are captured from labeled cells traversing a laser beam.[15] Impedance cytometers detect resistive pulses corresponding to cell volume displacement[23], while MFC employs magnetoresistive sensors to measure transient field perturbations from magnetically labeled targets[24,25]. These signals are segmented to detect, count, and classify individual events. However, all modalities suffer from a critical limitation—signal overlap due to coincident events, wherein two or more cells simultaneously enter the detection zone, generating superimposed signal patterns.[26-37]

Coincidence is an especially pervasive challenge in dense clinical samples (e.g., whole blood) or under high-throughput conditions.[38,39] Reported coincidence rates range from 5–15% under typical experimental conditions and can exceed 20% in undiluted or rapidly acquired samples.[30,35,37,39,40-42] In applications targeting rare cells, such as hematopoietic stem cells or minimal residual disease detection, even a small fraction of ambiguous or discarded events can have an outsized effect on sensitivity and diagnostic confidence.[3,43-46] Critically, standard strategies mitigate coincidences by exclusion—through dilution, lysis, sheath flow, or gating—rather than true resolution.[32,35,37,47-49] E.g., patient sample testing without prior knowledge of the target analyte concentration is challenging for point-of-care cartridges, requiring a dilution step or a large dynamic concentration range to cover all patient types. These mitigation approaches introduce delays, increase sample handling complexity, and discard potentially valuable data, which is especially problematic for low-abundance targets or time-critical measurements.

To overcome the limitations of exclusion-based or thresholding techniques, we present a hybrid signal reconstruction framework that directly disentangles coincident cell events into their constituent single-cell signals. Our approach integrates two complementary methods: (1) a fully convolutional neural network (FCN) trained on microbead-derived data to quantify cell coincidences using transfer learning and (2) a compressive sensing (CS) algorithm that reconstructs overlapping signals as sparse combinations of parameterized templates. Unlike conventional gating or thresholding methods, which discard ambiguous signals, and purely data-driven models that require retraining for each new cell type or condition, our framework uniquely combines model generalization with explicit signal interpretability. This solution enables end-to-end resolution and analysis of raw sensor data into biologically meaningful, interpretable single-cell signals—without requiring sample dilution, sheath flow, or manual exclusion criteria.

Trained on bead-derived signals, the FCN accurately classifies time-series events and generalizes to immunomagnetically labeled primary cells—including CD4 T cells and CD14 monocytes—without additional retraining. This transfer learning strategy enables rapid deployment across different sample types and clinical targets, significantly reducing the need for extensive, labor-intensive dataset-specific training. The CS component decomposes the signals by fitting them as sparse superpositions of adaptable templates—each capable of modeling peak asymmetry, amplitude, and variability—recovering key biophysical features such as cell size, velocity, and signal amplitude. In validation experiments on a custom MFC system, including optical tracking as ground truth, the combined approach correctly identifies and resolves coincident events with high accuracy (> 97%), demonstrating reliable recovery of otherwise lost single-cell data.

Although demonstrated with MFC, this framework generalizes to a wide range of sensing platforms that produce 1D time-resolved signals, including impedance cytometry, resistive pulse sensing, and nanopore platforms.[50] By reframing coincidence not as a nuisance to be excluded but as a rich signal source to be resolved[43,51,52], our work establishes a sensor-agnostic, clinically scalable solution to improve single-cell measurement fidelity, while enabling cell-interaction studies. This approach promises significant gains in diagnostic sensitivity, streamlined rare-cell detection workflows, and robust data recovery in high-speed or complex biological samples—extending the utility of high-throughput cytometry and biosensing in both research and translational applications.[47,53-55]

## 2. Results and Discussion

*2.1. Two-Stage Signal Disentanglement: FCN Classification and CS Reconstruction*

We developed a two-stage signal-processing framework, combining an FCN and CS to classify and reconstruct coincident cell events in MFC (Fig. 1). This approach empowers versatile and quantitative signal interpretation of complex biological samples at single-cell resolution—without requiring dilution optimization or extensive sample preprocessing.

The FCN is trained to estimate the number of overlapping cells or beads within each signal, while the CS module subsequently reconstructs individual signal components from the composite waveform. Our MFC platform uses immunomagnetically labeled cells or superparamagnetic beads, which are magnetically focused on the magnetic sensor unit. Single events produce a characteristic four-peak waveform defined by the spatial configuration of the sensor layout. In contrast, coincident events yield complex, multi-peak signals with non-uniform morphologies arising from spatiotemporal overlap (Fig. 1a,c).

Optical reference data is inaccessible for blood-based magnetic sensing due to the opacity of whole blood and the buried sensor surface. However, bead measurements in a buffer allow simultaneous optical and electrical acquisition, enabling precise annotation of bead counts per signal. These annotated bead datasets serve as training input for the FCN, which learns to infer the number of coincident beads from signal morphology alone (Fig. 1b).

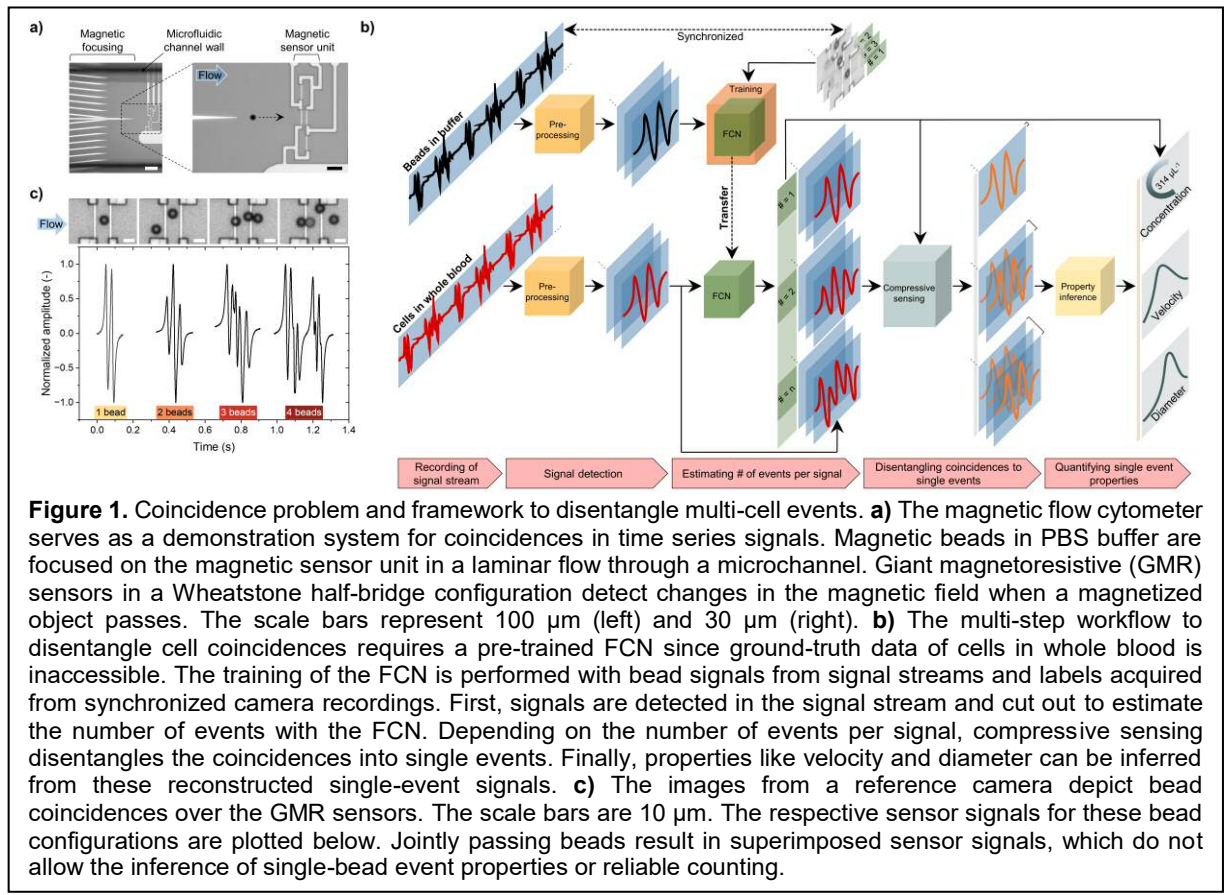

**Figure 1.** Coincidence problem and framework to disentangle multi-cell events. **a)** The magnetic flow cytometer serves as a demonstration system for coincidences in time series signals. Magnetic beads in PBS buffer are focused on the magnetic sensor unit in a laminar flow through a microchannel. Giant magnetoresistive (GMR) sensors in a Wheatstone half-bridge configuration detect changes in the magnetic field when a magnetized object passes. The scale bars represent 100 µm (left) and 30 µm (right). **b)** The multi-step workflow to disentangle cell coincidences requires a pre-trained FCN since ground-truth data of cells in whole blood is inaccessible. The training of the FCN is performed with bead signals from signal streams and labels acquired from synchronized camera recordings. First, signals are detected in the signal stream and cut out to estimate the number of events with the FCN. Depending on the number of events per signal, compressive sensing disentangles the coincidences into single events. Finally, properties like velocity and diameter can be inferred from these reconstructed single-event signals. **c)** The images from a reference camera depict bead coincidences over the GMR sensors. The scale bars are 10 µm. The respective sensor signals for these bead configurations are plotted below. Jointly passing beads result in superimposed sensor signals, which do not allow the inference of single-bead event properties or reliable counting.

Once trained, the FCN is transferred to cell data without retraining. This cross-domain deployment demonstrates the model's capacity for generalization and eliminates the need for laborious retraining across different cell types or patient samples. The predicted count from the FCN is then used to parameterize the CS algorithm, which reconstructs individual cell events for final single-cell property inference.

The recorded signal stream is first passed through a preprocessing module that detects transient signals using a threshold on the moving standard deviation, enabling robust, noise-adaptive segmentation. Detected signals are then fed into the FCN, which predicts the number of coincident cells. The predicted count for each signal is used to configure the CS algorithm that reconstructs individual event components for the composite waveform. Each event is modeled as a superposition of four Gaussian-shaped curves, resembling a single-cell event characterized by alternating polarity peaks. This parametrization allows flexible adaptation to variations in peak amplitude, shape, and asymmetry within each signal and event, accommodating biological heterogeneity across diverse sample types.

The resulting reconstructed single-cell signals sum precisely to the original input yet now offer a disentangled, individual-event resolution. From these disentangled events, key biophysical properties—including diameter, velocity, epitope density, and morphology—are extracted using established MFC signal analysis methods. Additionally, cell concentration is inferred directly from the FCN classification, enabling comprehensive, quantitative sample profiling.

### 2.2. FCN Performance and Generalization Across Bead Datasets

To evaluate the classification performance and cross-domain generalization, we trained FCNs on datasets from beads of varying diameters and relative magnetite content (e.g., 53%), reflecting differences in magnetic moment. These variations yield distinct signal amplitudes and morphologies (Fig. 2a), resembling the heterogeneity encountered in biological samples.

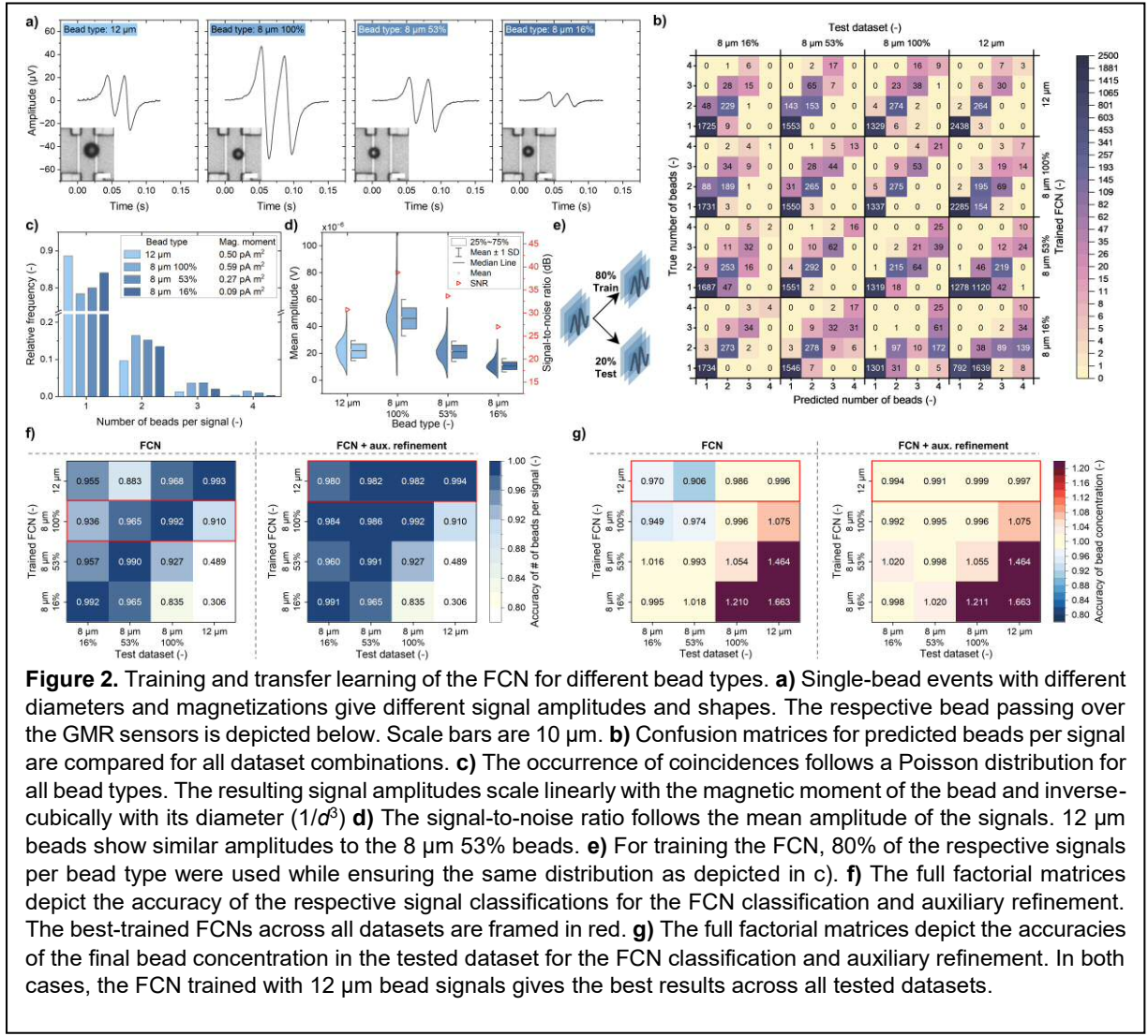

**Figure 2.** Training and transfer learning of the FCN for different bead types. **a)** Single-bead events with different diameters and magnetizations give different signal amplitudes and shapes. The respective bead passing over the GMR sensors is depicted below. Scale bars are 10 µm. **b)** Confusion matrices for predicted beads per signal are compared for all dataset combinations. **c)** The occurrence of coincidences follows a Poisson distribution for all bead types. The resulting signal amplitudes scale linearly with the magnetic moment of the bead and inverse-cubically with its diameter ($1/d^3$) **d)** The signal-to-noise ratio follows the mean amplitude of the signals. 12 µm beads show similar amplitudes to the 8 µm 53% beads. **e)** For training the FCN, 80% of the respective signals per bead type were used while ensuring the same distribution as depicted in c). **f)** The full factorial matrices depict the accuracy of the respective signal classifications for the FCN classification and auxiliary refinement. The best-trained FCNs across all datasets are framed in red. **g)** The full factorial matrices depict the accuracies of the final bead concentration in the tested dataset for the FCN classification and auxiliary refinement. In both cases, the FCN trained with 12 µm bead signals gives the best results across all tested datasets.

Bead-specific signal characteristics are governed by the magnetic moment and the geometric scaling of the induced magnetic field in the sensor plane. Although 12 µm beads possess higher magnetic moments than 8 µm 53% beads, their larger diameter reduces the net field change due to inverse-cubic scaling ($d^{-3}$), leading to comparable signal amplitudes (Fig. 2c–d). Signal-to-noise ratios (SNRs) followed a similar trend: highest for 8 µm 100% beads, followed by 8 µm 53%, 12 µm, and lowest for 8 µm 16% beads (Fig. 2d).

Each dataset was annotated using synchronized optical recordings and split into 80% training and 20% test sets, maintaining consistent bead-count distributions to avoid imbalance. Each FCN was trained on a single bead type and evaluated across all datasets using a full-factorial design (Fig. 2b,f). All FCNs achieved > 99% classification accuracy on their respective training datasets. However, generalization performance varied with the SNR and bead diameter, and models typically underestimated the number of beads per event (Fig. 2b).

The FCN trained on high-SNR data (8 µm 100% beads) demonstrated strong generalization, achieving 93.6–96.5% classification accuracy across lower-SNR datasets (Fig. 2f). In contrast, FCNs trained on low-SNR data, particularly the 8 µm 16% beads, performed poorly on high-SNR inputs, often overestimating bead numbers (Fig. 2b,f), and inflating concentration estimates by up to +21% (Fig. 2g).

Notably, the FCN trained on the 12 µm dataset generalized effectively across all smaller-diameter datasets (88.3–96.8% classification accuracy), whereas the inverse transfer yielded poor results with accuracies as low as 30.6% (Fig. 2f). Bead concentration estimation was also most accurate using the FCN trained on 12 µm beads (90.6–98.6%) (Fig. 2g).

To mitigate misclassifications, we implemented a conservative auxiliary refinement based on signal morphology. If the number of peaks exceeded the plausible count for the predicted class (e.g., > 4 peaks for a single-bead event), predictions were adjusted upward by one class. This refinement improved classification accuracy to > 98% and concentration estimations to > 99% with the 12 µm-trained FCN (Fig. 2f,g). Without refinement, the 8 µm 100% FCN yielded the best standalone generalization (< 9% error); with refinement, the 12 µm FCN achieved < 2% error across all datasets.

To interpret FCN decision-making, class activation mapping (CAM) was applied to the 12 µm-trained model (Fig. 3). For one- and two-bead events, the FCN focused on salient signal features such as peak shape and interpeak transitions across all datasets. As bead coincidence levels increased, attention increasingly shifted to noise in low-SNR datasets (e.g., 8 µm 16%), possibly interpreting signal length as a distinguished feature. In contrast, FCNs retained evident attention on meaningful signal features in high-SNR datasets, even for complex events.

Collectively, these results highlight the advantages of training on high-fidelity datasets. While classification accuracy is central for downstream coincidence disentangling *via* CS, concentration estimation depends directly on FCN outputs. FCNs trained on high-SNR or large-diameter bead signals demonstrated robust generalization across diverse input types. This transferability, especially when augmented with morphological refinement, lays a strong foundation for extending FCN-based classification to cell-derived signals with similar waveform complexity.

2.3. *High-Fidelity Signal Reconstruction Using Compressive Sensing*

Following FCN-based estimation of the number of coincident beads per signal, CS was applied to disentangle individual bead contributions from composite events. Each signal was modeled as a superposition of four parametrized Gaussian peaks per predicted bead, enabling adaptive fitting to variations in amplitude, asymmetry, and overall signal shape. The summed reconstruction, representing the total of all inferred single-bead events, was compared with the raw signal to assess reconstruction fidelity.

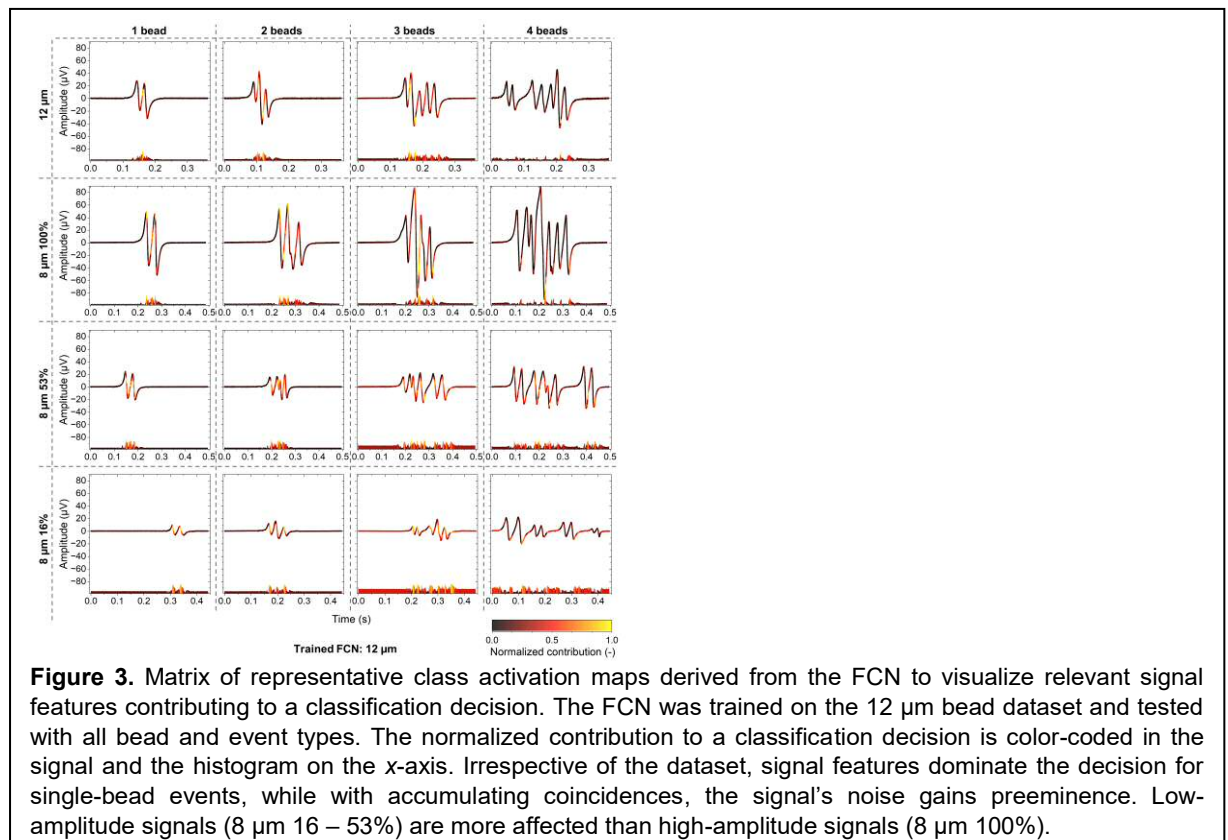

**Figure 3.** Matrix of representative class activation maps derived from the FCN to visualize relevant signal features contributing to a classification decision. The FCN was trained on the 12 µm bead dataset and tested with all bead and event types. The normalized contribution to a classification decision is color-coded in the signal and the histogram on the *x*-axis. Irrespective of the dataset, signal features dominate the decision for single-bead events, while with accumulating coincidences, the signal's noise gains preeminence. Low-amplitude signals (8 µm 16 – 53%) are more affected than high-amplitude signals (8 µm 100%).

For coincidence events, where ground truth for individual components is not directly accessible, reconstruction fidelity was evaluated by comparing the reconstructed signal with the corresponding raw waveform across bead types and coincidence levels (Fig. 4a–d). Cross-correlation between reconstructed and raw signals consistently exceeded 0.96, indicating high fidelity across all tested conditions. Notably, correlation coefficients often increased with the number of coinciding beads, underscoring the robustness of the CS framework in resolving increasingly complex signal overlaps. Among all tested conditions, the 8 µm beads with 100% magnetic content exhibited the highest consistency, with mean deviations in signal integrals and amplitudes ranging from –5.5% to +0.1% across all coincidence levels.

In contrast, beads with lower magnetic content (e.g., 8 µm 16%) exhibited greater reconstruction variability. Deviations in signal integral ranged from –0.1% to –7.6%, and in mean amplitude from –11.3% to +1.1%. These fluctuations increased with coincidences, indicating that lower SNR impairs reconstruction stability. The 12 µm beads also showed amplitude deviations up to –13.4% but with lower variability across coincidence levels, reflecting a more consistent negative offset. In general, CS-reconstructed signals slightly underestimated both amplitude and integral relative to the raw signal. Higher SNR was associated with reduced deviations and fluctuations, confirming that signal quality influences reconstruction fidelity, while different bead diameters introduced systematic offsets.

For isolated single-bead events, where reference values are available from established benchmark methods, reconstruction accuracy was assessed by directly comparing reconstructed signal properties with their raw counterparts. Hydrodynamic diameter, mean amplitude, and velocity were used to determine reconstruction accuracy at the single-event level. For all 8 µm beads, reconstructed diameters differed by +1.7% to +4.7% and velocity by +0.8% to +2.5%. The 12 µm beads showed negligible deviations in both diameter and velocity. However, mean amplitudes were consistently underestimated across all bead types, with deviations ranging from –10.4% to –5.2%, indicating mild but systematic signal attenuation during reconstruction.

Visual inspection of representative examples further supports these findings (Fig. 4f). Reconstructed waveforms closely matched the raw signals, even for complex 3- and 4-bead events, with only minor deviations in amplitude or peak structure. These results highlight the capability of the CS framework to accurately disentangle and reconstruct individual signal components from overlapping events. Given that cellular signals in whole blood are expected to have similarly low SNRs, the demonstrated reconstruction performance provides a solid foundation for reliable application to biologically complex samples.

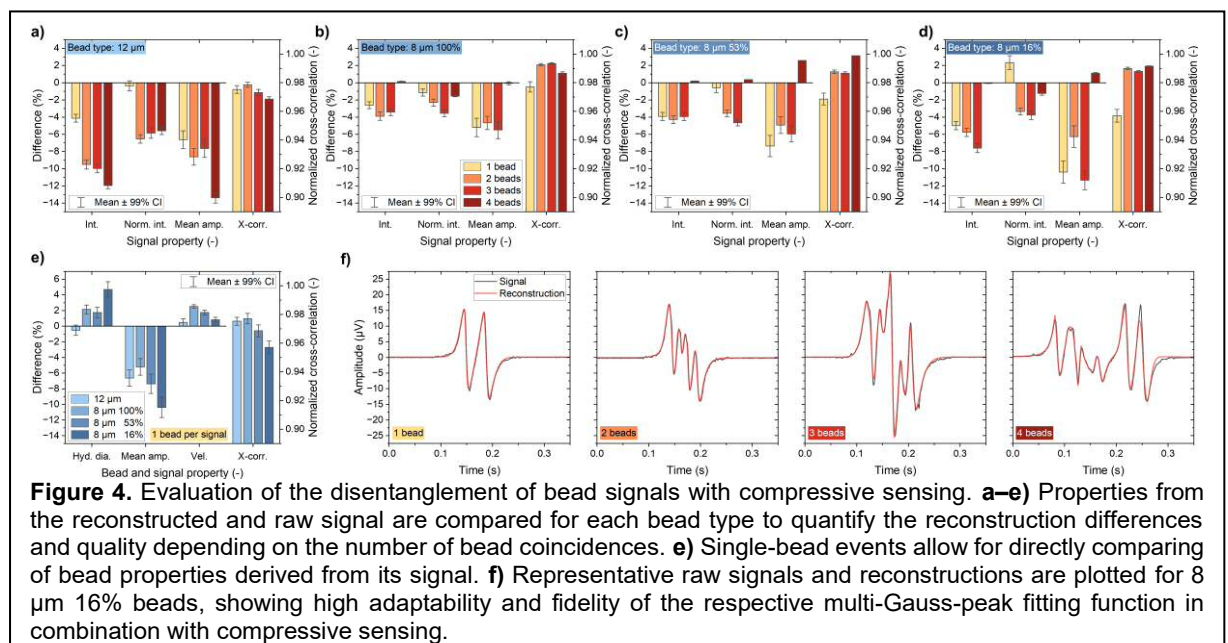

**Figure 4.** Evaluation of the disentanglement of bead signals with compressive sensing. **a–e)** Properties from the reconstructed and raw signal are compared for each bead type to quantify the reconstruction differences and quality depending on the number of bead coincidences. **e)** Single-bead events allow for directly comparing of bead properties derived from its signal. **f)** Representative raw signals and reconstructions are plotted for 8 µm 16% beads, showing high adaptability and fidelity of the respective multi-Gauss-peak fitting function in combination with compressive sensing.

## 2.4. Physical Property Recovery from Reconstructed Bead Signals

Key metrics—including hydrodynamic diameter, velocity, and amplitude—were compared to those from naturally occurring single-bead events to assess the reliability of bead property inference from reconstructed multi-bead events. For each bead type, single-bead properties were extracted from both isolated and disentangled signals, enabling the evaluation of consistency across coincidence levels (Fig. 5).

The hydrodynamic diameter was calculated from the peak-normalized integral and velocity from inter-peak distances. Scatter plots of diameter versus velocity, accompanied by histograms, revealed substantial overlap between 1- to 4-bead events (Fig. 5a). For all 8 µm beads, multi-bead reconstructions closely matched the shape of the single-bead distributions, with only minor shifts toward smaller diameters and slower velocities at higher coincidence levels. These deviations reflect slight amplitude underestimation and inter-peak spreading during reconstruction, particularly in 3- and 4-bead events. Systematic diameter overestimation was observed in lower-SNR beads, with median diameters of 8.7 µm, 9.1 µm, and 9.0 µm for 8 µm beads with 100%, 53%, and 16% magnetic content, respectively (Fig. 5d). In contrast, the 12 µm beads yielded a median of 12.0 µm, in agreement with manufacturer specifications and earlier results (Fig. 4e). These trends are consistent with previous observations (Fig. 4a–e), showing that reduced SNR leads to lower reconstructed amplitudes, which in turn yield higher normalized integrals and inflated diameter estimates.

Velocity reconstructions followed expected physical trends: 12 µm beads transited the sensor more rapidly (median: 740 mm s$^{-1}$) than 8 µm beads (range: 424–467 mm s$^{-1}$) (Fig. 5b). Velocity distributions for 8 µm beads were highly consistent across coincidence levels, with only slight shifts toward lower velocities in multi-bead events—likely due to minor errors in peak separation under more complex signal conditions (Fig. 5a). In contrast, 12 µm beads showed greater variability, with multi-bead signals tending to slower velocities.

A minor subpopulation with overestimated diameters, particularly among single-bead events, can be attributed to misestimated integration bounds (Fig. 5a; labeled as systematic error). These bounds are determined based on a metric that depends on antipodal peak spacing and amplitude and may occasionally include signal onset artifacts or noise, leading to artificially elevated integrals.

Amplitude reconstructions further validated the accuracy of the CS framework. Mean amplitudes from multi-bead reconstructions (Fig. 5c) were consistent with those from isolated single-bead events and the original amplitude profiles (cf. Fig. 2d). Even for the low-SNR 8 µm 16% beads, which approximate cellular signal conditions, amplitude distributions remained stable across 1- to 4-bead events. Median amplitudes were 9.4, 9.9, 9.9, and 10.9 µV, respectively (Fig. 5e). Outliers with elevated amplitudes were observed predominantly in the 4-bead subset, indicating modest reconstruction variability under complex overlap conditions. However, these rare deviations did not substantially impair overall reconstruction fidelity.

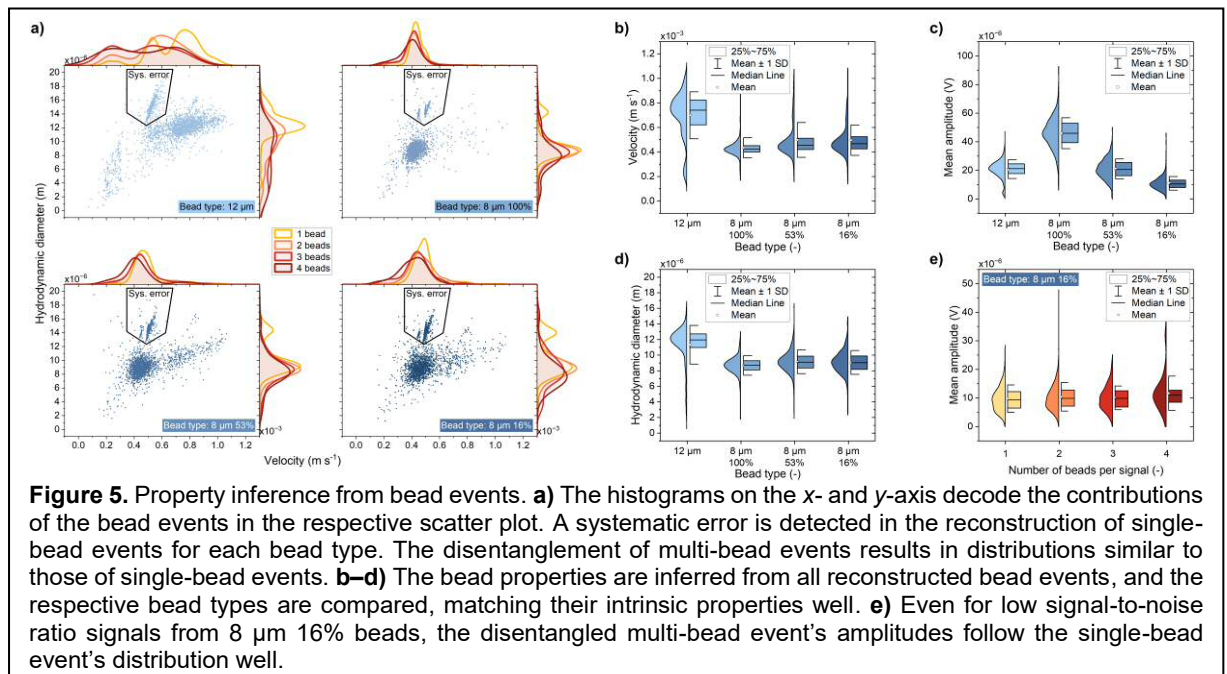

**Figure 5.** Property inference from bead events. **a)** The histograms on the *x*- and *y*-axis decode the contributions of the bead events in the respective scatter plot. A systematic error is detected in the reconstruction of single-bead events for each bead type. The disentanglement of multi-bead events results in distributions similar to those of single-bead events. **b–d)** The bead properties are inferred from all reconstructed bead events, and the respective bead types are compared, matching their intrinsic properties well. **e)** Even for low signal-to-noise ratio signals from 8 µm 16% beads, the disentangled multi-bead event's amplitudes follow the single-bead event's distribution well.

These findings demonstrate that the CS framework enables accurate recovery of key physical bead properties, even in highly overlapped signals. Reconstruction performance remained largely unaffected by signal complexity, SNR, or bead size. With appropriate gating strategies tailored to specific applications, target subpopulations can be extracted with high fidelity, reflecting actual physical distributions. These findings serve as a crucial step toward applying the framework to biological samples, where overlapping signals and the absence of ground truth complicate conventional analyses. The consistency between reconstructed and native single-bead properties underscores the framework's suitability for deployment in settings lacking ground truth, such as cellular analysis in whole blood.

*2.5. Transfer Learning to Biological Samples: From Beads to Cells*

To evaluate the applicability of the disentangling framework to biological samples, we applied the whole pipeline to immunomagnetically labeled CD4 T lymphocytes and CD14 monocytes in whole blood, measured *via* MFC. These cell types differ significantly in size, approx. 7 µm for CD4 and 13 µm for CD14 cells—thus providing a stringent benchmark for classification and signal reconstruction under physiologically relevant conditions.

Rather than training directly on cellular data, we leverage FCNs pre-trained on bead signals. Reference labels for CD4 and CD14 signals were assigned through expert inspection of raw signal traces without optical ground truth for coincident events. This design enables the evaluation of transfer learning from bead datasets, which can be generated rapidly and at scale using automated video-based recognition.

Cell signals were more heterogeneous, noisy, and challenging to classify than bead signals. The raw data exhibited baseline drift, abrupt signal jumps, and lower amplitudes, often below those of 8 µm 16% beads. Single-cell events from CD4 cells showed substantial variability in amplitude, peak spacing, and symmetry (Fig. 6a,c). These artifacts impaired classification performance, particularly for multi-cell signals. Without auxiliary refinement—a post-processing step originally developed for bead signals—the predicted cell counts were consistently underestimated, most prominently for 4-cell and CD14 signals (Fig. 6d).

Incorporating auxiliary refinement significantly improved classification accuracy. For CD14 signals, accuracy increased from 0.80–0.88 to 0.96 (Fig. 6e). The 8 µm 100% FCN achieved the highest accuracies across both datasets (0.83 for CD4 and 0.96 for CD14), though models trained on 12 µm and 8 µm 53% beads also performed comparably. Gains were smaller for CD4 cells, likely due to intrinsic noise and greater signal variability.

Cell concentration estimation further highlighted the benefit of refinement. Without post-processing, the 8 µm 53% FCN yielded concentration errors below 2.3% across both datasets. Refinement reduced this error to under 1.2% with the 8 µm 16% FCN (Fig. 6g). Across all models, the auxiliary step constrained the concentration error between –6.6% and +5.2%, with the most significant improvement again observed for CD14 samples. These results indicate that while the FCN predictions principally transfer from bead to cells, the refinement step effectively corrects systematic biases in classification and quantification.

Model interpretability was analyzed *via* CAMs (Fig. 6f). For 1- and 2-cell signals, peak amplitude, and inter-peak slope dominated the decision space, consistent with bead-based CAMs (cf. Fig. 3). Notably, in 3- and 4-cell signals, the influence of noise became markedly more pronounced, diminishing the contribution of physically meaningful features. This shift reflects the impact of reduced SNR on classification reliability while confirming that the transfer learning approach remains robust for simpler signal compositions, comprising more than 95% of all signals (Fig. 6a).

Although no single model emerged as optimal across all metrics, the 12 µm bead-trained FCN provided the best compromise between classification accuracy and concentration estimation. These findings support using bead datasets for pre-training, enabling scalable and generalizable application of signal disentangling frameworks to complex biological measurements.

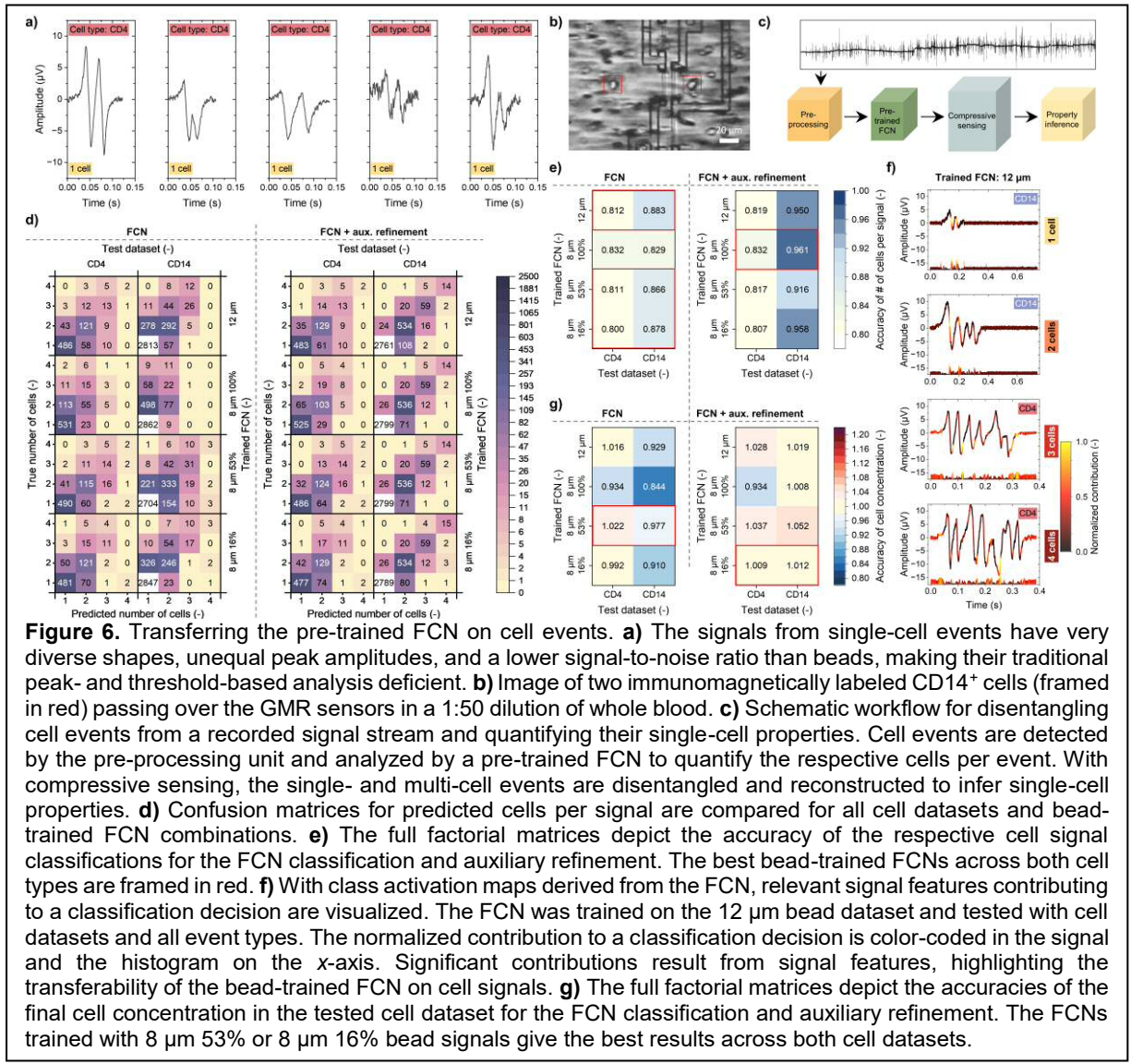

**Figure 6.** Transferring the pre-trained FCN on cell events. **a)** The signals from single-cell events have very diverse shapes, unequal peak amplitudes, and a lower signal-to-noise ratio than beads, making their traditional peak- and threshold-based analysis deficient. **b)** Image of two immunomagnetically labeled CD14+ cells (framed in red) passing over the GMR sensors in a 1:50 dilution of whole blood. **c)** Schematic workflow for disentangling cell events from a recorded signal stream and quantifying their single-cell properties. Cell events are detected by the pre-processing unit and analyzed by a pre-trained FCN to quantify the respective cells per event. With compressive sensing, the single- and multi-cell events are disentangled and reconstructed to infer single-cell properties. **d)** Confusion matrices for predicted cells per signal are compared for all cell datasets and bead-trained FCN combinations. **e)** The full factorial matrices depict the accuracy of the respective cell signal classifications for the FCN classification and auxiliary refinement. The best bead-trained FCNs across both cell types are framed in red. **f)** With class activation maps derived from the FCN, relevant signal features contributing to a classification decision are visualized. The FCN was trained on the 12 µm bead dataset and tested with cell datasets and all event types. The normalized contribution to a classification decision is color-coded in the signal and the histogram on the *x*-axis. Significant contributions result from signal features, highlighting the transferability of the bead-trained FCN on cell signals. **g)** The full factorial matrices depict the accuracies of the final cell concentration in the tested cell dataset for the FCN classification and auxiliary refinement. The FCNs trained with 8 µm 53% or 8 µm 16% bead signals give the best results across both cell datasets.

## 2.6. Validation Against Conventional State-Machine Detection

To benchmark our framework, we compared it against a conventional state-machine algorithm, which identifies events using amplitude and peak-spacing thresholds for subsequent property inference. This method assumes that each signal originates from a single cell and cannot distinguish between single- and multi-cell events. As such, it is fundamentally limited in its ability to analyze highly concentrated samples where coincident signals are prevalent.

Critically, our deep-transfer-learning-based classification substantially improved total cell recovery. FCNs trained on bead-derived datasets accurately predicted the number of CD4 and CD14 cells in whole blood samples, recovering 97% and 92% of total cell events, respectively. In contrast, the state-machine detected only 76% of CD4 and 72% of CD14 cells, counting 21 and 20 percentage points fewer cells, respectively (Fig. 7g). This disparity is particularly consequential for detecting rare cell populations or applications in clinical contexts where precise quantification is essential.

Signal coincidence, a major contributor to the state machine's poor performance, followed a Poisson distribution, as illustrated in Figure 7a. Between 19% and 27% of all measured signals represented overlaps from multiple cells. The state machine, unable to resolve such superimposed events, undercounts total cells and distorts the accuracy of downstream phenotypic assessments.

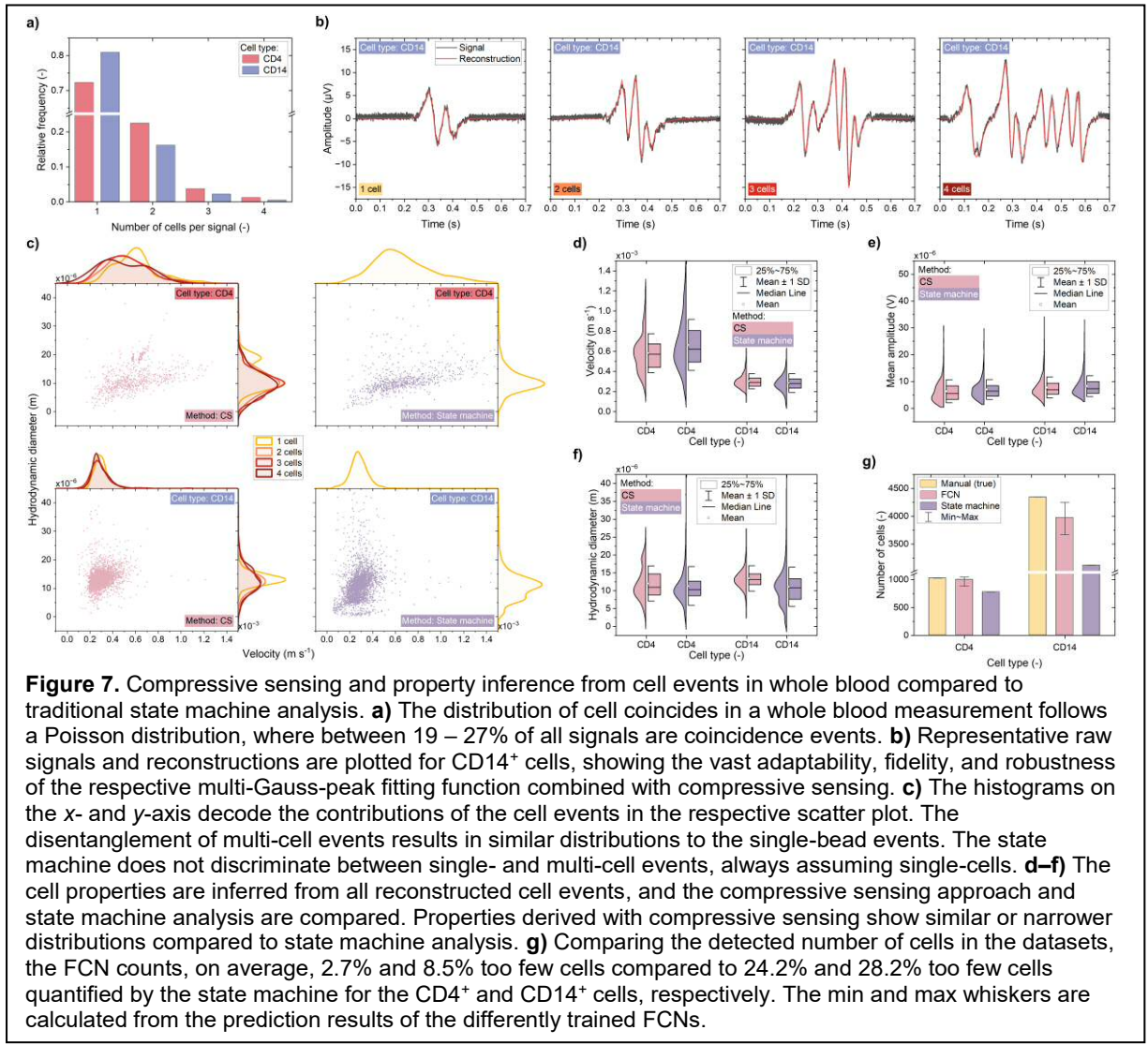

**Figure 7.** Compressive sensing and property inference from cell events in whole blood compared to traditional state machine analysis. **a)** The distribution of cell coincides in a whole blood measurement follows a Poisson distribution, where between 19 – 27% of all signals are coincidence events. **b)** Representative raw signals and reconstructions are plotted for CD14[+] cells, showing the vast adaptability, fidelity, and robustness of the respective multi-Gauss-peak fitting function combined with compressive sensing. **c)** The histograms on the *x*- and *y*-axis decode the contributions of the cell events in the respective scatter plot. The disentanglement of multi-cell events results in similar distributions to the single-bead events. The state machine does not discriminate between single- and multi-cell events, always assuming single-cells. **d–f)** The cell properties are inferred from all reconstructed cell events, and the compressive sensing approach and state machine analysis are compared. Properties derived with compressive sensing show similar or narrower distributions compared to state machine analysis. **g)** Comparing the detected number of cells in the datasets, the FCN counts, on average, 2.7% and 8.5% too few cells compared to 24.2% and 28.2% too few cells quantified by the state machine for the CD4[+] and CD14[+] cells, respectively. The min and max whiskers are calculated from the prediction results of the differently trained FCNs.

To overcome this, we applied CS to deconvolve cell signals into their constituent single-cell events. Due to their size disparity and biological relevance, CD4 and CD14 cells were chosen as representative immune cell populations. Measurements were performed on whole blood in a clinically relevant workflow using MFC.

Cell signals are inherently more heterogeneous than bead signals, exhibiting greater variability in amplitude, asymmetric shapes, irregular peak spacing, and elevated baseline noise. Despite these challenges, CS-based signal reconstruction proved highly robust and adaptable (Fig. 7b). The multi-Gaussian peak fitting model effectively captured morphological variability across a various SNRs. As with beads, minor amplitude underestimation persisted, but this did not impair downstream property inference.

Following signal disentanglement, single-cell properties—including hydrodynamic diameter, velocity, and amplitude—were inferred and benchmarked against the state-machine-derived results. Across both CD4 and CD14 cell populations, CS reconstruction consistently yielded more coherent and less variable distributions. In detail, standard deviations in diameter were reduced by 10% for CD4 and 35% for CD14 cells and in velocity by 24% and 21%, respectively (Fig. 7d,f). The increased spread in the state-machine-derived distributions stems from its failure to resolve coincident signals, resulting in composite and misleading single-cell profiles.

Further, CD4 diameter distributions reconstructed *via* CS revealed a secondary peak, also observed in single-bead data, indicative of a known systematic signal artifact (Fig. 7c). In contrast, state-machine-derived distributions, particularly for CD14 cell diameters, were broader and more irregular, reflecting the misattribution of multi-cell signals as single-cell events. Notably, CS-derived distributions remained consistent across different coincident levels, mirroring results from bead experiments and underscoring the robustness of the reconstruction algorithm (Fig. 7c).

Amplitude distributions were more comprehensive under CS analysis (Fig. 7e). This reflects the method's retention of low-amplitude peaks, typical in multi-cell events, which the state machine's fixed-threshold approach disregards. Specifically, the state machine considers only the first four peaks above a defined amplitude cutoff, systematically truncating signals and introducing bias towards higher amplitude subsets. This results in clipped histograms and diminished sensitivity to signal heterogeneity.

Median property values further underscore the fidelity of the CS-based approach. For CD4 cells, diameters were 11.0 µm (CS) vs. 10.3 µm (state machine), while for CD14 cells, they were 13.1 µm vs. 10.8 µm, respectively. The state-machine method thus diminished inter-population differences, whereas CS preserved biologically relevant distinctions. Observed velocity differences—CD4 cells appearing faster than CD14—were attributed to different flow rates during acquisition (25 µL min$^{-1}$ vs. 20 µL min$^{-1}$) and not intrinsic cellular properties (Fig. 7d).

Our hybrid framework—combining FCN-based classification with CS-driven signal disentanglement—enables accurate and robust analysis of complex cell signals in noisy environments. It restores accurate cell counts, resolves overlapping events, and infers single-cell biophysical properties with high fidelity. These capabilities are essential for advancing quantitative flow-based diagnostics and cell profiling in heterogeneous and clinically relevant sample types.

## 3. Conclusions

Accurate identification and reconstruction of coincident cell events remain a critical challenge in single-cell biosensing, where signal overlap compromises sensitivity, reliability, and throughput, particularly in complex clinical samples. We present a robust reconstruction framework integrating an FCN with CS to resolve coincident events from one-dimensional sensor data. By leveraging transfer learning from bead-derived signals, the FCN generalizes effectively to immunomagnetically labeled primary cells in minimally processed whole blood, eliminating labor-intensive retraining. The CS component further enables the high-fidelity decomposition of overlapping signals, recovering distinct biophysical features for each individual cell. Benchmarking against conventional state-machine approaches demonstrates superior accuracy, robustness to noise, and inference of cell properties across diverse acquisition conditions.

The framework is sensor-agnostic and adaptable to various time-resolved single-cell technologies, including magnetic flow cytometry, impedance cytometry, resistive pulse sensing, and nanopore-based analyzers. Its integration of explainable AI elements—such as class activation maps and parameterized signal templates—adds interpretability crucial for clinical diagnostics and regulatory workflows.

Future directions will focus on expanding to diverse cell types and sensing modalities, reducing data requirements *via* self-supervised pretraining, and validating in translational settings such as rare-cell detection, immune profiling, and therapeutic monitoring. Altogether, this work lays a solid foundation for next-generation single-cell sensing platforms that are automated, scalable, and capable of resolving coincident events in high-throughput, clinically relevant environments, broadening the utility of cytometric and biosensing tools in biomedical research and precision diagnostics.

## 4. Experimental Section

### 4.1. Magnetic flow cytometer setup

The design and operational principle of the MFC have been previously described.[56,57] In brief, a structured silicon die featuring giant magnetoresistive (GMR) sensing elements (2 µm width, 30 µm length) configured in a Wheatstone half-bridge was positioned above a permanent magnet. Chevron-shaped magnetic rails of nickel-iron alloy were patterned onto the die to magnetically guide and focus particles onto the sensing elements. The microfluidic chip consisted of a PDMS channel (700 µm width

× 150 µm height) through which samples were pumped at controlled rates (20–30 µL min$^{-1}$) using a pulsation-free syringe pump.[58] The change in resistance of the GMR sensors, induced by passing magnetized objects, was sensed as a voltage change, sampled at 10 kS s$^{-1}$ with 16-bit resolution. Optical monitoring was performed using a 20× objective lens and a CMOS camera to verify flow conditions and generate ground truth data for bead classification.

### 4.2. Sample preparation and bead measurements

Superparamagnetic polystyrene-magnetite microbeads (micromer-M, micromod Partikeltechnologie GmbH) were suspended in DPBS supplemented with 0.5% polysorbate 20 and 0.5% BSA. Bead suspensions were measured at a flow rate of 30 µL min$^{-1}$. Immunomagnetically labeled cells were diluted 1:30 in DPBS and measured at 20 or 25 µL min$^{-1}$.

### 4.3. Immunomagnetic cell labeling

Peripheral blood was collected from healthy donors under informed consent and stabilized with EDTA (S-Monovette EDTA K3E, 9 mL, Sarstedt GmbH). For CD14$^+$ monocyte labeling, a commercial kit (StraightFrom Whole Blood CD14 MicroBeads, Miltenyi Biotec) was used, comprising magnetic nanoparticles functionalized with anti-CD14 antibodies. Whole blood was mixed with the nanoparticles at a 2:1 ratio and incubated at room temperature on a rotary shaker for 5–6 hours.

CD4$^+$ T lymphocytes were labeled using dextran-coated magnetic nanoparticles and anti-CD4 antibodies (EasySep Human CD4 Positive Selection Kit II, STEMCELL Technologies). The antibodies and nanoparticles were pre-incubated at a 1:1 ratio overnight at 3–8 °C before mixing with whole blood at a 1:9 ratio and incubated for 30 minutes on a rotary shaker.

### 4.4. Signal detection and preprocessing

Signals were extracted using a dynamic noise-adaptive thresholding approach. Events were identified when the moving standard deviation exceeded a threshold relative to the baseline noise, estimated from the minimum value across moving standard deviation windows. Each signal was cropped with a buffer and subjected to quality control criteria, including standard deviation thresholds, signal length, zero crossings, and peak counts. Parameter settings are detailed in the Supporting Information.

### 4.5. Fully convolutional network (FCN) training

Signal classification was performed using a one-dimensional FCN adapted from Wang et al. and Baur et al.[59,60] The model consisted of three convolutional blocks (kernel sizes: 1×16, 1×10, 1×6) with batch normalization and ReLU activation, followed by global average pooling, a dense layer, and softmax classification. Signals were padded with noise (estimated from the initial 50 time points) to ensure uniform input length. The training was conducted for up to 500 epochs using negative log-likelihood loss and the Adam optimizer with a learning rate of 0.001.[61] Datasets were split 80:20 for training and testing, with stratification to preserve bead count distributions.

### 4.6. Compressive sensing (CS) implementation

CS exploits the sparsity of the reconstruction vector to recover overlapping signal components. In this approach, adapted from Baur et al., the sensing matrix is defined as the superposition of parameterized templates, each resembling a single-cell signal.[60] These templates are initially drawn from a compact template library and serve as starting points for parameter optimization.

The CS reconstruction solves a minimization problem to identify the optimal combination of template parameters that best reconstruct the measured signal. Each template consists of the superposition of four independent Gaussian peaks, each described by four parameters (amplitude , shape factor , center , and standard deviation ):

$$\quad (1)$$

This formulation provides greater flexibility in modeling variable peak shapes, asymmetries, and amplitudes than earlier approaches (see Supporting Information Fig. S3).[60] Accurately capturing the signal heterogeneity is critical for robust and precise signal decomposition, particularly of overlapping or distorted events. The minimization process adjusts the template parameters to best match the observed signal in a sparse representation space. This allows the model to resolve multi-cell events and recover individual signal components with high fidelity, thereby enabling subsequent inference of single-cell biophysical properties.

*4.7. Property inference and benchmarking*

The FCN-CS framework and the state-machine algorithm used comparable methods to extract signal properties. Velocities were calculated from inter-peak distances, while hydrodynamic diameters were estimated using a third-order polynomial calibration curve relating the peak-normalized signal integral to diameter.[22,56]

For state-machine-based detection, amplitude thresholds were set to 3.0 µV for $CD4^+$ cells and 4.2 µV for $CD14^+$ cells, corresponding to approximately 4× and 7× the baseline noise standard deviation. Lower thresholds increased false-positive rates due to noise artifacts.

**Declaration of competing interest**

The authors declare no competing interests.

**CRediT authorship contribution statement**

**Moritz Leuthner**: Conceptualization, Methodology, Software, Validation, Formal Analysis, Investigation, Data Curation, Writing – Original Draft, Writing – Review & Editing, Visualization. **Rafael Vorländer**: Software, Formal Analysis, Investigation, **Oliver Hayden**: Resources, Writing – Review & Editing, Project Administration, Funding Acquisition, Supervision.

**Data availability**

The data supporting the conclusions of this article will be made available upon reasonable request.

**Acknowledgments**

The authors would like to thank Anika Kwiatkowski (Technical University of Munich) for her support with wire bonding the sensor chips and Michael Wack (Ludwig-Maximilians-Universität München) for facilitating the use of their vibrating sample magnetometer for quantifying the beads' magnetic moment.

This study (406/20 S-EB) was approved by Ethikkommission an der Technischen Hochschule München, approved on 19 October 2020. Informed consent was obtained from each study participant.

**Supporting Information**

Supporting information to this article can be found online.

**References**

[1] Stuart, T., Satija, R., Integrative single-cell analysis, 2019, Nature Reviews Genetics, 20, 257-72, 10.1038/s41576-019-0093-7
[2] Barnett, D., Walker, B., Landay, A., Denny, T. N., CD4 immunophenotyping in HIV infection, Nature Reviews Microbiology, 2008, S7-15, 10.1038/nrmicro1998

[3]  Jaso, J. M., Wang, S. A., Jorgensen, J. L., Lin, P., Multi-color flow cytometric immunophenotyping for detection of minimal residual disease in AML: past, present and future, 2014, Bone Marrow Transplantation, 49, 1129-38, 10.1038/bmt.2014.99

[4]  Chattopadhyay, P. K., Gierahn, T. M., Roederer, M., Love, J. C., Single-cell technologies for monitoring immune systems, 2014, Nature Immunology, 15, 2, 128-35, 10.1038/ni.2796

[5]  Chen, X., Cherian, S., Acute Myeloid Leukemia Immunophenotyping by Flow Cytometric Analysis, 2017, Clin. Lab. Med., 37, 753-69, 10.1016/j.cll.2017.07.003

[6]  Wojas-Krawczyk, K., Kalinka, E., Grnda, A., Krawczyk, P., Milanowski, J., Beyond PD-L1 Markers for Lung Cancer Immunotherapy, 2019, Int. J. Mol. Sci., 20, 1915, 10.3390/ijms20081915

[7]  Liu, E., Tong, Y., Dotti, G., et al., Cord blood NK cells engineered to express IL-15 and a CD19-targeted CAR show long-term persistence and potent antitumor activity, 2018, Leukemia, 32, 520-31, 10.1038/leu.2017.226

[8]  Mazziotta, F., Martin, L. E., Egan, D. N., et al., A phase I/II trial of WT1-specific TVR gene therapy for patients with acute myeloid leukemia and active disease post-allogeneic hematopoietic cell transplantation: skewing towards NK-like phenotype impairs T cell function and persistence, 2025, Nat. Comm., 16, 5214, 10.1038/s41467-025-60394-0

[9]  Melenhorst, J. J., Chen, G. M., Wang, M., et al., Decade-long leukaemia remissions with persistence of $CD4^+$ CAR T cells, 2022, Nature, 602, 503-9, 10.1038/s41586-021-04390-6

[10] Nolan, J. P., Yang, L., The flow of cytometry into systems biology, 2007, Breifings in Functional Genomics and Proteomics, 6, 2, 81-90, 10.1093/bfgp/elm011

[11] Fritzsch, F. S. O., Dusny, C., Frick, O., Schmid, A., Single-Cell Analysis in Biotechnology, Systems Biology, and Biocatalysis, 2012, Annu. Rev. Chem. Biomol. Eng., 3, 129-55, 10.1146/annurev-chembioeng-062011-081056

[12] Suhail Y., Cain, M. P., Vanaja, K., Kurywchak, P. A., Levchenko, A., Kalluri, R., Kshitiz, Systems Biology of Cancer Metastasis, 2019, Cell Systems, 9, 109-27, 10.1016/j.cels.2019.07.003

[13] Kuckuck, F. W., Edwards, B. S., Sklar, L. A., High Throughput Flow Cytometry, 2001, Cytometry, 44, 83-90, 10.1002/1097-0320(20010501)44:1<83::AID-CYTO1085>3.0.CO;2-O

[14] McKinnon, K. M., Flow Cytometry: An Overview, 2018, Current Protocols in Immunology, 120, 5.1.1-11, 10.1002/cpim.40

[15] Shapiro, H. M., 2003, Practical flow cytometry, 4th ed., Wiley (Hoboken), ISBN: 0-471-41125-6

[16] Bacheschi, D. T., Polsky, W., Kobos, Z., Yosinski, S., Menze, L., Chen, J., Reed, M. A., Overcoming the sensitivity vs. throughput tradeoff in Coulter counters: A novel side counter design, 2020, Biosens. Bioelectr., 168, 112507, 10.1016/j.bios.2020.112507

[17] Cheung, C. K., Di Berardino, M., Schade-Kampmann, G., Hebeisen, M., Pierzchalski, A., Bocsi, J., Mittag, A., Tárnok, A., Microfluidic Impedance-Based Flow Cytometry, 2010, Cytometry Part A, 77A, 648-66, 10.1002/cyto.a.20910

[18] Holmes, D., Pettigrew, D., Reccius, C. H., et al., Leukocyte analysis and differentiation using high speed microfluidic single cell impedance cytometry, 2009, Lab Chip, 9, 2881-89, 10.1039/b910053a

[19] Sun, T., Morgan, H., Single-cell microfluidic impedance cytometry: a review, 2010, Microfluid. Nanofluid., 8, 423-43, 10.1007/s10404-010-0580-9

[20] Helou, M., Reisbeck, M., Tedde, S. F., et al., Time-of-flight magnetic flow cytometry in whole blood with integrated sample preparation, 2013, Lab Chip, 13, 1035-38, 10.1039/c3lc41310a

[21] Issadore, D., Chung, J., Shao, et al., Ultrasensitive Clinical Enumeration of Rare Cells ex Vivo Using Micro-Hall Detector, 2012, Sci. Transl. Med., 4, 141ra92, 10.1126.scitranslmed.3003747

[22] Reisbeck, M., Richter, L., Helou, M. J., et al., 2018, Biosens. BIoelectr., 109, 98-108, 10.1016/j.bios.2018.02.046

[23] Hoffmann, R. A., Johnson, T. S., Britt, W. B., Flow Cytometric Electronic Direct Current Volume and Radiofrequency Impedance Measurements of Single Cells and Particles, 1981, Cytometry, 1, 6, 377-84, 10.1002/cyto.990010605

[24] Boser, B. E., Murali, P., Flow Cytometer-on-a-Chip, 2014, IEEE Biomedical Circuits and Systems Conference (BioCAS) Proceedings, 480-83, 10.1109/BioCAS.2014.6981767

[25] Loureiro, J., Andrade, P. Z., Cardoso, S., da Silva, C. L., Cabral, J. M., Freitas, P. P., Magnetoresistive chip cytometer, 2011, Lab Chip, 11, 2255, 10.1039/c01c00324g

[26] Burel, J. G., Pomaznoy, M., Lindestam Arlehamn, C. S., et al., Circulating T cell-monocyte complexes are markers of immune pertubations, 2019, eLife, 8, e46045, 10.7554/eLife.46045

[27] Burel, J. G., Pomaznoy, M., Lindestam Arlehamn, C. S., Seumois, G., Vijayanand, P., Sette, A., Peters, B., The Challenge of Distinguishing Cell-Cell Complexes from Singlet Cells in Non-Imaging Flow Cytometry and Single-Cell Sorting, 202, Cytometry Part A, 97A, 1127-35, 10.1002/cyto.a.24027

[28] Caselli, F., De Ninno, A., Reale, R., Businaro, L., Bisegna, P., A Bayesian Approach for Coincidence Resolution in Microfluidic Impedance Cytometry, 2021, IEEE Transactions on Biomedical Engineering, 68, 1, 340-9, 10.1109/TBME.2020.2995364

[29] Hassan, U., Bashir, R., Coincidence detection of heterogeneous cell populations from whole blood with coplanar electrodes in a microfluidic impedance cytometer, 2014, Lab Chip, 14, 4370, 10.1039/c4lc00879k

[30] Higgins, A. Z., Karlsson, J. O. M., Coincidence Error During Measurement of Cellular Osmotic Properties by the Electrical Sensing Zone Method, 2008, CryoLetters, 29, 6, 447-61, https://www.ingentaconnect.com/contentone/cryo/cryo/2008/00000029/00000006/art00001

[31] Jie, C., Ahmed, R., Hamad, A. R. A., Expression of unique gene signature distinguishes $TCRαβ^+/BCR^+$ dual expressers from $CD3^+CD14^+$ doublets, 2022, Cytometry Part A, 101, 283-9, 10.1002/cyto.a.24542

[32] Keij, J. F., van Rotterdam, A., Groenewegen, A. C., Stokdijk, W., Visser, J. M. V., Coincidence in High-Speed Flow Cytometry: Models and Measurements, 1991, Cytometry, 12, 398-404, 10.1002/cyto.990120504

[33] Kellman, M. R., Rivest, F. R., Pechacek, A., Sohn, L. L., Lustig, M., Node-Pore Coded Coincidence Correction: Coulter Counters, Code Design, and Sparse Deconvolution, 2018, IEEE Sensors Journal, 18, 8, 3068-79, 10.1109/JSEN.2018.2805865

[34] Pisani, J. F., Thomson, G. H., Coincidence errors in automatic particle counters, 1971, J. Phys. E: Sci. Instrum., 4, 5, 359-61, 10.1088/0022-3735/4/5/006

[35] Rico, L. G., Bardina, J., Bistué-Rovira, À., Salvia, R., Ward, M. D., Bradford, J. A., Petriz, J., Accurate identification of cell doublet profiles: Comparison of light scattering with fluorescence measurement techniques, 2023, Cytometry Part A, 103, 447-54, 10.1002/cyto.a.24690

[36] Wu, T. Y., Murashima, Y., Sakuari, H., Iida, K., A bilateral comparison of particle number concentration standards via calibration of an optical particle counter for number concentration up to ~1000 cm$^{-3}$, 2022, Measurement, 189, 110446, 10.1016/j.measurement.2021.110446

[37] Wynn, E. J. W., Hounslow, M. J., Coincidence correction for electrical-zone (Coulter-counter) particle size analysers, 1997, Powder Technology, 93, 163-75, 10.1016/S0032-5910(97)03267-1

[38] Edmundson, I. C., Coincidence Error in Coulter Counter Particle Size Analysis, 1966, Nature, 212, 1450-2, 10.1038/2121450b0

[39] Petriz, J., Bradford, J. A., Ward, M. D., No lyse no wash flow cytometry for maximizing minimal sample preparation, 2018, Methods, 134-135, 149-63, 10.1016/j.ymeth.2017.12.012

[40] Chen, Y., Hu, Y., Hu, H., Advances and challenges in platelet counting: evolving from traditional microscopy to modern flow cytometry, 2025, J. Lab. Med., 49, 1, 2-13, 10.1515/labmed-2024-0135

[41] Hassan, U., Reddy, B. Jr., Damhorst, G., Sonoiki, O., Ghonge, T., Yang, C., Bashir, R., A microfluidic biochip for complete blood cell counts at the point-of-care, 2015, Technology, 3, 4, 201-13, 10.1142/S2339547815500090

[42] Peng, L., Wang, W., Bai, L., Performance evaluation of the Z2 coulter counter for WBC and RBC counting, 2007, Int. Jnl. Lab. Hem., 29, 361-8, 10.1111/j.1751-553X.2007.00868.x

[43] Bendall, S. C., Diamonds in the doublets, 2020, Nature Biotechnology, 38, 559-61, 10.1038/s41587-020-0511-6

[44] Böhmer, R. M., Bandala-Sanchez, E., Harrison, L. C., Forward Light Scatter Is a Simple Measure of T-Cell Activation an Proliferation but Is Not Universally Suited for Doublet Discrimination, 2011, Cytometry Part A, 79A, 646-52, 10.1002/cyto.a21096

[45] Fornas, O., Garcia, J., Petriz, J., Flow cytometry counting of CD34$^+$ cells in whole blood, 2000, Nature Medicine, 6, 7, 833-6, 10.1038/77571

[46] Greve, B., Beller, C., Cassens, U., Sibrowski, W., Severin, E., Göhde, W., High-Grade Loss of Leukocytes and Hematopoietic Progenitor Cells Caused by Erythrocyte-Lysing Procedures for Flow Cytometric Analysis, 2003, J. Hematotherapy and Stem Cell Research, 12, 321-30, 10.1089/152581603322023052

[47] Huffman, P. A., Arkoosh, M. R., Characteristics of Peripheral Blood Cells from Rainbow Trout Evaluated by Particle Counter, Image Analysis, and Hemocytometric Techniques, 1997, J. Aquatic Animal Health, 9, 4, 239-48, 10.1577/1548-8667(1997)009<0239:COPBCF>2.3.CO;2

[48] Kundernatsch, R. F., Letsch, A., Stachelscheid, H., Volk, H.-D., Scheibenbogen, C., 2013, Cytometry Part A, 83A, 173-6, 10.1002/cyto.a.22247

[49] Reitz, S., Kummrow, A., Kammel, M., Neukammer, J., Determination of micro-litre volumes with high accuracy for flow cytometric blood cell counting, 2010, Meas. Sci. Technol., 21, 074006, 10.1088/0957-0233/21/7/074006

[50] Yusko, E. C., Bruhn, B. R., Eggenberger, O.M., et al., Real-time shape approximation and fingerprinting of single proteins using a nanopore, 2017, Nature Nanotechnology, 12, 360-7, 10.1038/NNano.2016.267

[51] Giladi, A., Cohen, M., Medaglia, C., et al., Dissecting cellular crosstalk by sequencing physically interacting cells, 2020, Nature Biotechnology, 38, 629-37, 10.1038/s41587-020-0442-2

[52] Wang, Y., Wang, Z., Yang, J., et al., Deciphering Membrane-Protein Interactions and High-Throughput Antigen Identification with Cell Doublets, 2024, Adv. Sci., 11, 2305750, 10.1002/advs.202305750

[53] Krips, R., Furst, M., Stochastic Properties of Coincidence-Detector Neural Cells, 2009, Neural Computation, 21, 2524-53, 10.1162/neco.2009.07-07-563

[54] Stuart, G. J., Häusser, M., Dendritic coincidence detection of EPSPs and action potentials, 2001, Nature Neuroscience, 4, 1, 63-71, 10.1038/82910

[55] Wen, C., Dematties, D., Zhang, S.-L., A Guide to signal Processing Algorithms for Nanopore Sensors, 2021, ACS Sens., 6, 3536-55, 10.1021/acssensors.1c01618

[56] Leuthner, M., Helou, M., Reisbeck, M., Hayden, O., Advancing magnetic flow cytometry to quantitative epitope analysis in high hematocrit conditions for point-of-care testing, 2025, Biosens. Bioelectr., 268, 116867, 10.1016/j.bios.2024.116867

[57] Leuthner, M., Reisbeck, M., Helou, M., Hayden, O., Towards a Point-of-Care Test of CD4$^+$ T Lymphocyte Concentrations for Immune Status Monitoring with Magnetic Flow Cytometry, 2024, Micromachines, 15, 520, 10.3390/mi15040520

[58] Leuthner, M., Hayden, O., Grease the gears: how lubrication of syringe pumps impacts microfluidic flow precision, 2024, Lab Chip, 24, 56-62, 10.1039/d3lc00698k

[59] Wang, Z., Yan, W., Oates, T., Time Series Classification from Scratch with Deep Neural Networks: A Strong Baseline, 2017, International Joint Conference on Neural Networks (IJCNN), 1578-85, 10.1109/IJCNN.2017.7966039

[60] Baur, M., Reisbeck, M., Hayden, O., Utschick, W., Joint Particle Detection and Analysis by a CNN and Adaptive Norm Minimization Approach, 2022, IEEE Transactions on Biomedical Engineering, 69, 8, 2468-79, 10.1109/TBME.2022.3147701

[61] Kingma, D., Ba, J., Adam: A Method for Stochastic Optimization, 2014, arXiv, 10.48550/arXiv.1412.6980

# Supporting Information

# Disentangling coincident cell events using deep transfer learning and compressive sensing

*Moritz Leuthner\*, Rafael Vorländer, Oliver Hayden\**

M. Leuthner[1,2,3], R. Vorländer[1,3], O. Hayden[1,2,3]
[1] Heinz-Nixdorf-Chair of Biomedical Electronics, School of Computation, Information and Technology, Technical University of Munich, Einsteinstraße 25, 81675 Munich, Germany
[2] Munich Institute of Biomedical Engineering, School of Computation, Information and Technology, Technical University of Munich, Boltzmannstraße 11, 85748 Garching b. Munich, Germany
[3] Central Institute for Translational Cancer Research (TranslaTUM), University Hospital Rechts der Isar, Technical University of Munich, Einsteinstraße 25, 81675 Munich, Germany
E-mail: moritz.leuthner@tum.de; oliver.hayden@tum.de

## 1. Signal detection and preprocessing – STEP 1+5

**Table S1.** Parameters for the signal detection and the preprocessing.

| Description | Parameter | Bead or cell type | | | | | |
|---|---|---|---|---|---|---|---|
| | | 12 µm | 8 µm 100% | 8 µm 53% | 8 µm 16% | CD4[+] cells | CD14[+] cells |
| Signal scaling factor | scaling | 1 E6 | 1 E6 | 1 E6 | 1 E6 | 1 E6 | 1 E6 |
| Moving standard deviation window size | std_intervall | 200 | 200 | 200 | 200 | 200 | 200 |
| Base noise average window size | std_thrcalc_intervall | 5000 | 5000 | 5000 | 5000 | 5000 | 5000 |
| Base noise average window step | std_thrcalc_step | 1000 | 1000 | 1000 | 1000 | 1000 | 1000 |
| Base noise multiplier | std_thr_mult | 5 | 5 | 2 | 2 | 3 | 2 |
| Buffer size | buffer | 100 | 100 | 100 | 100 | 150 | 300 |
| Min. signal length | min_length | 400 | 400 | 400 | 400 | 600 | 1400 |
| Max. signal length | max_length | 8000 | 8000 | 8000 | 8000 | 8000 | 8000 |
| Min. moving standard deviation | min_amp | 1.836 E-6 | 1.649 E-6 | 0.542 E-6 | 0.542 E-6 | 1.402 E-6 | 0.020 E-6 |
| Max. moving standard deviation | max_amp | 1836 E-6 | 1649 E-6 | 542 E-6 | 542 E-6 | 1402 E-6 | 20 E-6 |
| Max. bias jump | jump_goodwill | 0.1 | 0.1 | 0.1 | 0.1 | 0.4 | 0.4 |

| | | | | | | |
|---|---|---|---|---|---|---|
| Min. zero crossings | min_zero_cross | 2 | 2 | 2 | 2 | 3 | 3 |
| Min. number of peaks | min_peaks | 2 | 2 | 2 | 2 | 2 | 2 |

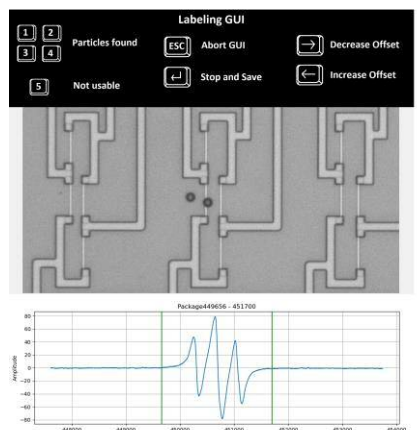

**Figure S1.** Graphical user interface (GUI) for semi-automated signal labeling.

## 2. FCN training – STEP 2

**Table S2.** Parameters for the FCN training.

| | | Bead or cell type | | | | | |
|---|---|---|---|---|---|---|---|
| **Description** | **Parameter** | 12 µm | 8 µm 100% | 8 µm 53% | 8 µm 16% | CD4[+] cells | CD14[+] cells |
| Dataset duplication factor | DUPLICATION_FACTOR | 1 | 1 | 1 | 1 | n.a. | n.a. |
| Shuffle dataset before split | SHUFFLE | False | False | False | False | n.a. | n.a. |
| Smooth signals | filter | False | False | False | False | n.a. | n.a. |
| Smoothing method | gaussian_filter(sigma) | n.a. | n.a. | n.a. | n.a. | n.a. | n.a. |
| Normalize signals | norm | False | False | False | False | n.a. | n.a. |
| Train / test dataset ratio | TRAIN_DATA_RATIO | 0.8 | 0.8 | 0.8 | 0.8 | n.a. | n.a. |
| Shuffle train dataset | shuffle | True | True | True | True | n.a. | n.a. |
| Shuffle test dataset | shuffle | False | False | False | False | n.a. | n.a. |
| Learning rate | LEARNING_RATE | 0.0001 | 0.0001 | 0.0001 | 0.0001 | n.a. | n.a. |
| Batch size | BATCH_SIZE | 64 | 64 | 64 | 64 | n.a. | n.a. |
| Number of trainings | num_trainings | 5 | 5 | 5 | 5 | n.a. | n.a. |

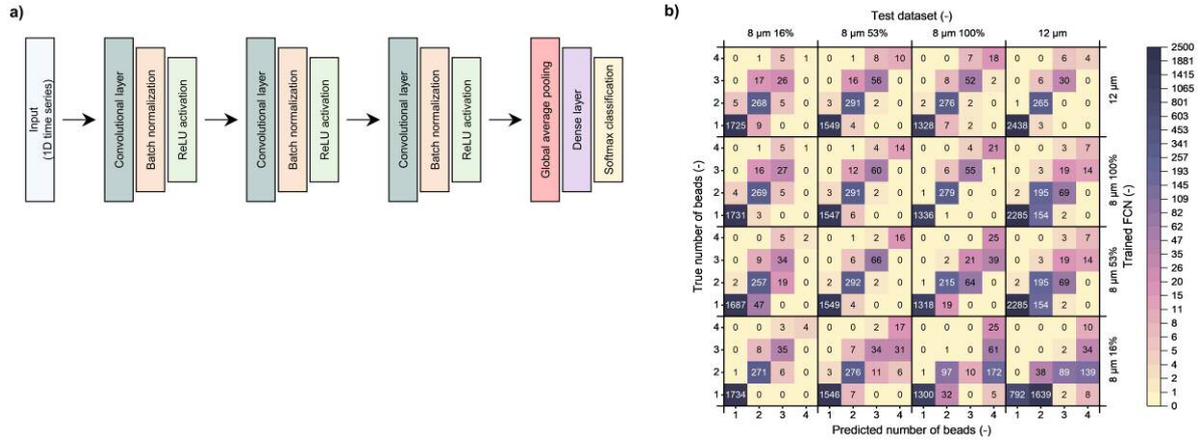

**Figure S2. a)** FCN architecture. **b)** Confusion matrices for predicted beads per signal with auxiliary refinement are compared for all dataset combinations.

## 3. FCN transfer testing – STEP 6

**Table S3.** Parameters for the FCN transfer testing.

| Description | Parameter | Bead or cell type | | | | | |
|---|---|---|---|---|---|---|---|
| | | 12 μm | 8 μm 100% | 8 μm 53% | 8 μm 16% | CD4+ cells | CD14+ cells |
| Shuffle dataset | SHUFFLE; shuffle | False | False | False | False | False | False |
| Smooth signals | filter | False | False | False | False | True | True |
| Smoothing method | gaussian_filter(sigma) | n.a. | n.a. | n.a. | n.a. | 35 | 35 |
| Normalize signals | norm | False | False | False | False | True | True |

## 4. Disentanglement with CS – STEP 3

**Table S4.** Parameters for the signal disentanglement with CS.

| Description | Parameter | Bead or cell type | | | | | |
|---|---|---|---|---|---|---|---|
| | | 12 μm | 8 μm 100% | 8 μm 53% | 8 μm 16% | CD4+ cells | CD14+ cells |
| GMR distance | DISTANCE_GMR | 14 E-6 | 14 E-6 | 14 E-6 | 14 E-6 | 18 E-6 | 16 E-6 |
| Noise for signal padding | add_noise_pad | True | True | True | True | True | True |
| Smooth signals | filter | True | True | True | True | True | True |
| Smoothing method | savgol_filter( window_length, polyorder) | (25, 3) | (25, 3) | (25, 3) | (25, 3) | (25, 3) | (25, 3) |
| Normalize signals | norm | False | False | False | False | False | False |
| Tolerances for least squares depending on label | ftols | E-2, E-6, E-6, E-6 | E-2, E-6, E-6, E-6 | E-2, E-6, E-6, E-6 | E-2, E-6, E-6, E-6 | E-2, E-6, E-6, E-6 | E-2, E-6, E-6, E-6 |
| Time out for least squares fitting | timeout | 45 | 45 | 45 | 45 | 45 | 45 |
| Epochs for Adam optimizer | epochs | 2000 | 2000 | 2000 | 2000 | 2000 | 2000 |

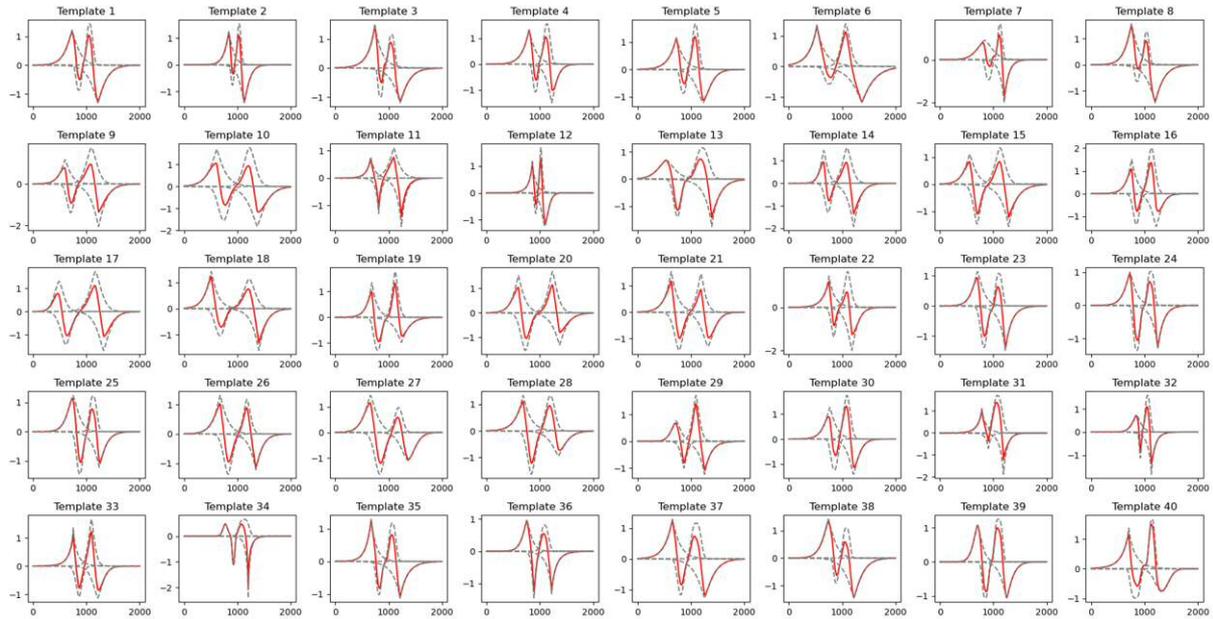

**Figure S3.** Templates used for signal disentanglement with CS. The dashed lines represent the Gaussian curves with their sum depicted in red. For disentangling signals with CS, the Gaussian parametrization is used.

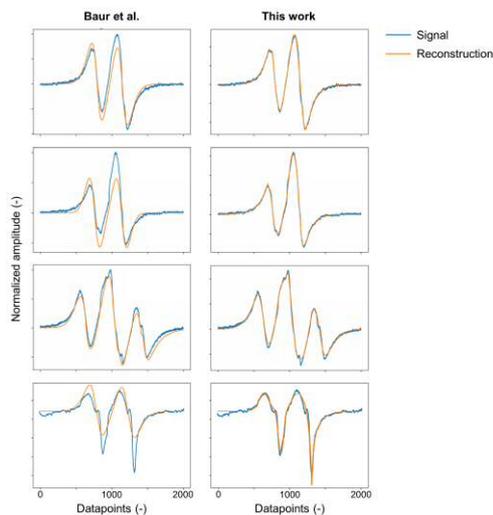

**Figure S4.** Qualitative comparison between the CS approach from Baur et al. and this work. Baur et al. reconstructed a single-bead signal with two superimposed derivatives of a Gaussian function. This approach does not allow for individual fitting of peak amplitudes and widths, resulting in less reconstruction fidelity compared to our approach.

## References

Baur, M., Reisbeck, M., Hayden, O., Utschick, W., Joint Particle Detection and Analysis by a CNN and Adaptive Norm Minimization Approach, 2022, IEEE Transactions on Biomedical Engineering, 69, 8, 2468-79, 10.1109/TBME.2022.3147701